# On the Roles of Graphene Oxide Doping for Enhanced Supercurrent in MgB$_2$

W. K. Yeoh,[a,b,c,*] X. Y. Cui,[a,b] B. Gault,[d] K. S. B. De Silva,[c] X. Xu,[c] H. W. Liu,[a] H. W. Yen,[a,b] D. Wong,[e] P. Bao,[e] D. J. Larson,[f] I. Martin,[f] W. X. Li,[c,g] R. K. Zheng,[e] X. L. Wang,[c] S. X. Dou,[c,*] and S. P. Ringer[a,b,]

[a] School of Aerospace, Mechanical and Mechatronic Engineering, The University of Sydney, New South Wales 2006, Australia.

[b] Australian Centre for Microscopy & Microanalysis, University of Sydney, Sydney, New South Wales 2006, Australia.

[c] Institute for Superconducting & Electronic Materials, University of Wollongong, Innovation Campus, Squires Way, North Wollongong, New South Wales 2500, Australia.

[d] Department of Materials, University of Oxford, Parks Road, Oxford OX1 3PH, UK.

[e] School of Physics, The University of Sydney, New South Wales 2006, Australia.

[f] CAMECA Instruments Inc., 5500 Nobel Drive, Madison, WI 53711, USA.

[g] Solar Energy Technologies, School of Computing, Engineering & Mathematics, University of Western Sydney, Penrith, New South Wales 2751, Australia.

Due to its graphene-like properties after oxygen reduction, incorporating graphene oxide (GO) sheets into correlated-electron materials offer a new pathway for tailoring their properties. Fabricating GO nanocomposites with polycrystalline MgB$_2$ superconductor leads to an order of magnitude enhancement of the supercurrent at 5 Kelvin /8 Tesla and 20 Kelvin /4 Tesla. Herein, we introduce a novel experimental approach to overcome the formidable challenge of performing quantitative microscopy and microanalysis of such composites, so as to unveil how GO doping influences the structure and hence the material properties. Atom probe microscopy and electron microscopy were used to directly image the GO within the MgB$_2$, and we combined these data with computational simulations to derive the property-enhancing mechanisms. Our results reveal synergetic effects of GO, namely, via localized atomic (carbon and oxygen) doping as well as texturing of the crystals, which provide both inter- and intra-granular flux pinning. This study opens up new insights into how low-dimensional nanostructures can be integrated into composites to modify the overall properties, using a methodology amenable to a wide range of applications.

*Address correspondence to
waikong.yeoh@sydney.edu.au
shi@uow.edu.au

**Introduction**

Understanding and controlling the local microstructure of electron-correlated systems so as to tailor their functional properties is a burgeoning area of modern materials engineering.[1] This is especially true for superconductors, where the ability to tune the localized ordering has led to remarkable progress in both research into the fundamentals of superconductivity[2] and research aimed at applications.[3] The critical current density ($J_c$) is usually the key property that determines whether superconductors are plausible candidates for technological applications.[4,5] In general, defects are required to pin the magnetic flux lines so that large currents can flow through the material, even in the presence of high magnetic fields. The possibility of engineering effective pinning centers on the nanoscale has opened up new routes for optimizing superconductors with enhanced vortex pinning.[6-8]

Among all the efforts, chemical engineering with various carbon sources has been confirmed to be the most effective tactic to enhance the pinning centers.[9-12] Our group has previously shown the extraordinary effect on the $J_c$ of $MgB_2$ of low-dimensional functional doping with carbon via carbon nanotubes (CNTs)[13] and via graphene.[14] It remains a great experimental challenge, however, to adapt these approaches to more general application in bulk samples. In particular, there is great difficulty in quantifying the dopant distribution, and our lack of understanding of the dynamic interactions between dopants has seriously hindered progress in optimizing $J_c$. For example, the effect of oxygen, which often occurs as a tramp, or unintended, impurity element, on $J_c$ is still under debate.[15-17] The incorporation of oxygen usually results in the formation of MgO precipitate particles, and this is frequently cited as the limiting factor for $J_c$ because of the concomitant degradation in intergranular connectivity.[15] It was reported that nanoscale coherent precipitates of $Mg(B,O)_2$[16] and oxygen-substituted $MgB_{2-x}O_x$[17] can form in the $MgB_2$ matrix, and that these phases serve as highly effective flux pinning centers, thereby *increase* $J_c$. These inconsistencies in the reported effects of oxygen on $J_c$ expose the clear gap in our experimental capacity to correlate the oxygen atom distribution with material functionality.

Clearly, an imaging technique capable of directly addressing the atomic-scale microstructure is highly desirable. High-resolution transmission electron microscopy (HRTEM) is able to image single dopant atoms under very specific conditions, although chemical mapping of low-dimensional nanomaterials containing light elements remains extremely challenging. Recent developments in atom probe microscopy

(APM) have greatly advanced atomic resolution characterization,[18-20] including precise compositional mapping of semiconductor nanowires,[21] core-shell nanostructures,[22] and complex organic-inorganic interfaces.[23] Recently, we used this approach to reveal the dopant distribution in an iron-based superconductor, enabling detailed correlations between the atomic scale microstructure and the superconductivity.[24]

Graphene oxide (GO) is a two-dimensional nanostructured combination of carbon and oxygen functional groups. This material is attractive for a wide range of technological applications due to its tunable chemical and atomic structures. In particular, chemical tuning of GO platelets via chemical reduction is a growing area of research. One possible route to exploiting these functionalities is to use GO to create composite materials. Here, we investigate the effect of GO additions to $MgB_2$ on $J_c$.[25] By using APM, we are able to visualize individual GO platelets and map the distribution of GO in the $MgB_2$ bulk. APM enables direct observations of the preferred orientation of GO platelets, which promotes $MgB_2$ grain growth. Localized oxygen from GO reacts with the $MgB_2$ matrix to form nanoscale precipitates that yield an enhancement in $J_c$ beyond that obtained with CNTs or graphene.[13, 14] This produces a synergetic effect of localized atomic (carbon and oxygen) doping and texturing of the crystals that significantly enhances both low and high field $J_c$. This is in contrast with conventional processing techniques that focus on chemical tailoring and which seem to have reached a limit in terms of enhancement of the current density.[26]

**Results and discussion**

All samples were prepared via the diffusion method, which is known to achieve high density $MgB_2$ crystals and to minimize the formation of large MgO particles.[25] During the heat-treatment process, liquid Mg was infiltrated into premixed B and GO pellets and reacted to form $MgB_2$, as depicted in Figure 1(a). Figure 1(b) shows the field dependence of $J_c(B)$ curves at 5 and 20 K for graphene- and GO-doped samples, as well as for a pure $MgB_2$ control sample. It should be noted that both the GO and the graphene doped samples have the same nominal carbon doping level (0.5 wt.%) for the purposes of a true comparison. The GO-containing materials yielded the highest $J_c$ enhancement over a wide range of applied magnetic field at both 5 and 20 K, with $J_c$ values reaching $1 \times 10^4$ A/cm$^2$ at 5 K/8 T and 20 K/4 T. This is a full order of magnitude higher than for the (control) undoped sample. These results suggest that GO doped $MgB_2$ is a very promising composite material, not only for high field applications, but also for technological applications at 20 K.

In a related study, graphite-doping of MgB$_2$ was shown to improve both the irreversibility field ($H_{irr}$) and the $J_c$ as a result of carbon substitution for boron, changing the intrinsic properties such as the upper critical field ($H_{c2}$).[27] As a result, the improvement of $J_c$ in carbon-doped samples can be related to the enhancement of $H_{c2}$. In the GO-doped samples studied here, the normalized $H_{irr}$ improved gradually with increasing doping level, but the normalized $H_{c2}$ was largely unchanged beyond 1 wt.% GO, as shown in Figure 1(d). The unchanged normalized $H_{c2}$ suggests that the degree of carbon substitution is insignificant and does not induce increases in $J_c$ in our GO-doped samples. This is in agreement with a previous study of graphene-doped samples, where the carbon substitution effect was shown to be weak due to the strong carbon-carbon bonding of graphene.[28] Because the $J_c$ of GO-doped MgB$_2$ is higher than that of the graphene-doped sample, and $J_c$ increases with increasing GO-doping level,[25] we propose that a different flux pinning mechanism is at work.

Figure 2(a) presents an APM reconstruction of GO-doped MgB$_2$ superconductor, generated from a tomographic dataset of 7 million atoms: Mg, B, C, and O atoms are displayed, represented by dots that are colored so as to indicate their chemical identities. Importantly, the 1:2 ratio of Mg to B is preserved, indicating the accuracy of the analysis. In the three-dimensional (3D) tomographic reconstruction, the isoconcentration surfaces encompass regions of high C and O concentration and thus highlight the locations of individual GO platelets. It is obvious from visual inspection that the GO platelets are uniformly dispersed in the MgB$_2$ matrix, and most of the GO platelets appear to be perpendicular to the analysis direction, which corresponds to platelets parallel to the surface of the original MgB$_2$ sample. A profile obtained from these surfaces confirms the high local excess of C and O, which corresponds to GO sheets, either isolated or in stacks, as shown in the regions of interest (ROI) 1 and 2. Segregation of C and O was observed towards the bottom of the dataset (ROI 3), and this corresponds to a grain boundary, as indicated by a change in atomic density that is similar to those observed in other ceramics.[29]

The aligned GO platelets in the MgB$_2$ matrix serve as an interconnected network for textured MgB$_2$ grains, as demonstrated by our analysis of grain orientation, where the microstructural effects of GO doping on MgB$_2$ grain growth were clearly evident. The transmission Kikuchi diffraction (TKD) technique recently developed in our laboratory was used to analyze the distribution of grain orientations.[30] Figure 3(a) provides examples of the band contrast and inverse pole figure images derived from the TKD map after noise removal. There is a strong correspondence between the different colored regions

(representing different orientations) in the inverse pole figures and the band contrast image. Both the GO-doped and the undoped samples contain high-angle grain boundaries with > 10° misorientation with respect to adjacent grains, denoted by the black lines, while low-angle grain boundaries, denoted by the red lines, exhibited only ~5-10° misorientation. It is also apparent that there are regions where the grains share very similar and rather low (< 5°) misorientations, as shown in Figure 3(a). A quantitative analysis revealed that the GO-doped sample possessed a higher fraction of low-angle grain boundaries (5-10°) and a much smaller average grain size of 57.6 nm, compared to the average grain size in the control sample of 128.8 nm. Figure 3(b) compares the degree of preferred orientation of GO doped and pure $MgB_2$ in the {001} pole the $x_o$ and $y_o$ directions. The intensity distribution in the {001} pole figure suggests that *c*-axis alignment is the preferred orientation in both doped and undoped samples, and this probably originates from the fabrication technique. The tendency towards a single orientation, however, (marked as I) in the GO-doped samples compared to the two orientations (marked as I and II) in the undoped samples is clear evidence of a textured structure in the GO-doped sample. This implies that incorporation of GO-sheets not only promotes low-angle grain boundaries, but also constrains the grain growth of $MgB_2$ to specific orientations. Similar observations were made with respect to CNT doping, where high-aspect-ratio functional carbon sheets were shown to promote the growth of $MgB_2$ grains along preferred orientations.[31]

The different ratio of carbon to oxygen observed for the different stacks of sheets, as shown in Figure 2(b), strongly supports our hypothesis that oxygen and carbon could be removed from the surface of the GO platelets during thermal annealing.[32] This also suggests that oxygen can be introduced into the $MgB_2$ matrix, providing a localized source of oxygen doping, via the reduction of the GO. Taking into account that the GO precursor initially possesses 34% oxygen and exhibits an oxygen reduction efficiency of 60%[32], we estimate that the effective oxygen concentration incorporated into the $MgB_2$ matrix could be up to 1% for a GO doping level of 2%. Singh *et al*. showed that oxygen concentrations as low as 0.65 % were sufficient to produce $MgB_2$ films with high $J_c$.[33] In fact, it was suggested that the distribution of oxygen in the $MgB_2$ matrix plays a decisive role in the $J_c$ performance.

To understand the role and the spatial distribution of O in the microstructure, we calculated the formation energy[34] and the entropy of formation of various bulk phases containing oxygen via first principles density functional theory (DFT), using the generalized gradient approximation (see Supplementary Information Figure S1). The calculated heat entropy values of oxide based compounds such as MgO (1.01 eV) and $B_2O_3$ (0.91 eV) are much higher than for pure $MgB_2$ (0.145 eV), suggesting that oxygen atoms have a strong tendency to dissolve into the $MgB_2$ matrix as a solid solution. This behavior can be further understood by comparing the electronegativity values of the constituent elements: Mg (1.31), B (2.04), and O (3.44). Under the present B-rich conditions, atomic O exists competitively either in substitutional ($O_B$) or interstitial form ($O_i$), since the formation energies are close, with $O_i$ being slightly favored (2.17 eV versus 2.84 eV). While substitutional $O_B$ demonstrates a preference to be distributed uniformly within the $MgB_2$ crystal matrix, we note that interstitial $O_i$ is highly mobile[35] and forms a layered structure, with a single-triangle layer structure representing oxygen clusters ($MgB_2O_{0.25}$) or a double-triangle layer of oxygen clusters ($MgB_2O_{0.5}$). Both possess P31M space group symmetry, as shown in Figure 4(a). Under both O-rich and B-rich conditions, the calculated formation energies for single-triangle and double-triangle layer $O_i$ phases are lower than that for the single $O_i$. We further used TEM to examine the GO-doped sample and confirmed that the $O_i$ structure appeared adjacent to GO sheet within the $MgB_2$ matrix, as shown in Figure 4(b), and this was verified using fast Fourier transform (FFT) techniques. Although we were unable to distinguish these structures due to the resolution of the FFT, the presence of the interstitial phase was detected in X-ray diffraction as (101) with a peak at 41°, which is the major peak of precipitated $MgB_2O_{0.25}$ or $MgB_2O_{0.5}$. Detailed crystallographic information on the proposed $Mg(B,O)_2$ phases can be found in Figure S2 in the Supplementary Information.

Another significant observation from Figure 4(b) is that the $Mg(B,O)_2$ phase is surrounded by nanoparticles, which were identified as MgO. Conventional nucleation and growth processes normally produce much larger grain sizes of MgO (50-200 nm). The dispersion of nanoscale (~5-7 nm) MgO precipitates in Figure 4(b) indicates that their formation may occur via a spinodal-like phase separation. The microstructural evolution of Mg–B alloy was reported earlier, and compositional fluctuations and/or the presence of metastable phases have been reported as the major drivers for this sort of phase transformation.[36, 37] By comparing the heat entropy values of these ordered $Mg(B,O)_2$ phases with those of MgO and $B_2O_3$, it is evident that $Mg(B,O)_2$ is metastable and prone to decompose. This finding is in agreement with Liao et al., who reported the transformation of $Mg(BO)_2$ phase into MgO.[16] Mobile oxygen atoms reduced from the GO tend to form ordered phases when their concentration reaches a critical value, and microstructural modulation follows. During the cooling process, the solubility of oxygen decreases[16], forcing the oxygen to cluster locally, even if the average excess oxygen content of the sample is small. A miscibility gap between the oxygen concentrations drives the system towards phase separate, as exemplified by the $La_2CuO_{4+\delta}$ system.[38] Indeed, phase separation is a rather universal phenomenon affecting these doped systems – especially when dopants are present as interstitial solid solutions.[39, 40] As a result, the GO-doped samples demonstrate the formation of fine granular MgO precipitates that can be attributed to the phase separation associated with the excess oxygen, which, apparently, is a characteristic of the interstitial state of the dopant. These results also demonstrate that the reduction of GO occurs at around 500 °C[41], so that the localized oxygen could easily be incorporated into the $MgB_2$ matrix, which was also formed at this temperature. Therefore, we conclude that localized oxygen originating from the GO-doping behaves differently from the conventional O-treated $MgB_2$ fabricated with substantial oxygen-level, which exhibits a continuous of reactions that yield products which include $B_2O_3$, $MgB_4$, $MgB_6$, or $MgB_{12}$.[42, 43]

It is important to understand how the changes in the microstructure affect the pinning mechanisms both within a single grain (intragranular properties) and across grain boundaries (intergranular properties). According to Rowell[15], the change in resistivity accompanied by a change in temperature from 300 K to 40 K ($\Delta\rho$) can serve as a measure of the intergranular connectivity in $MgB_2$, while the residual resistivity [$\rho_0 = \rho(40\ K)$] can be corrected to determine the intragranular resistivity values. Figure 5 shows the values of $\Delta\rho$ (intergranular resistivity) and the corrected $\rho_0$ (intragranular resistivity) as a function of GO concentration. A decrease in $\Delta\rho$ with increasing GO doping indicates better connectivity between $MgB_2$ grains. The reduction of intergranular resistivity with increasing of GO-doping level is presumably a result of the presence of reduced GO sheets between the grains, as observed using APM. It is expected that the reduced GO has graphene-like electrical and thermal properties[44], and so improves the conductivity between grains in $MgB_2$. In un-textured polycrystalline $MgB_2$, the tendency towards randomly oriented grains results in anisotropic properties under an applied magnetic field. Specifically, $MgB_2$ grains with their *c*-axes oriented parallel to the magnetic field become normal conductors at relatively low fields, compared to grains oriented parallel to the *ab*-plane.[45] Thus, the number of superconducting grains decreases with increasing magnetic field, and finally, the current pathways become percolative, insofar as a continuous superconducting current path is no longer available.[45] This is responsible for the rapid decrease in $J_c$. As a result, the textured grains and a high fraction of low-angle grain boundaries oppose the percolation effect and promote long-range current pathways, especially at high magnetic fields. These factors have a strong effect on the superconducting transport properties across the boundaries and produce the intergranular effect referred to here.

In oxygen-doped $MgB_2$ films, a linear increase in the corrected $\rho_0$ with oxygen content is attributed to scattering by intragranular oxygen defects.[33] Consequently, the high $J_c$ in the GO-doped samples correlates well with the increased intragranular resistivity, and is due to the formation of nanoprecipitates within the grains (intragranular effect). This produces a synergetic effect, whereby the incorporation of GO induces both inter- and intra-granular pinning. More importantly, with the recent proposals for 'second generation' $MgB_2$ conductors[46], further optimization of $J_c$ seems possible by incorporating the present microstructural design principles. Our results are also analogous to the features observed in oxide-based doping, including $Dy_2O_3$[47], $Y_2O_3$[48], $Ho_3O_2$[49], and other inorganic doping[50], where in-situ reactions between the oxide dopant and the precursor powders provide localized oxygen and consequently lead to the formation of nanoscale MgO and boron-rich precipitate phases. Consequently, $H_{irr}$ is significantly enhanced by intragranular nanoscale precipitates; however, $H_{c2}$ is not significantly affected.[49, 50] This is indeed in contrast to carbon-doping, where the improvement in $J_c$ is closely related to an enhancement of $H_{c2}$. This is because substitutional C impurities have lower energies of formation than interstitial C,[51, 52] and the majority of C impurities in $MgB_2$ are substitutional, inducing a significant enhancement of $H_{c2}$.

**Experimental**

**Synthesis.** The samples were synthesized by the diffusion reaction method, based on the infiltration of magnesium (99.9%, −325 mesh) into premixed boron (99.999%, 0.2–2.4 μm) and graphene oxide. Firstly, the boron and GO were mixed and pressed into 13 mm diameter pellets. The pellets were then sintered at 800°C for 10 hours with 20% excess of nominal Mg in high purity argon gas. The sample was fabricated according to the weight ratio of $MgB_2$: GO equals to 100: $x$; where $x$ = 0, 0.5, 1.0, 1.5 and 2.0 wt.% of GO. Such a formation mechanism produces a dense structured $MgB_2$ with much less MgO impurity, which is difficult to achieve by the conventional powder-in-tube (PIT) process. GO was prepared and purified according to the modified Hummers method. The details of GO fabrication can be found in [25]. Graphite and graphene doped $MgB_2$ were also prepared under the same conditions for comparison purposes.

**Analysis.** The phase and crystal structure of all the samples were investigated using a MAC Science MX03 diffractometer with Cu Kα radiation. The magnetization was measured by a Physical Properties Measurement System (PPMS, Quantum Design) up to 13 T. Magnetic $J_c$ values were determined from the magnetization hysteresis loops using the appropriate critical state model. Bar-shaped samples with a size of 3 × 1.5 × 1 mm³ were cut for magnetic measurements. Electron microscopy studies were carried out using a JEOL 2200FS instrument, with a 200 kV field emission gun and in-column omega type energy filter, equipped with electron energy loss spectroscopy (EELS). Our atomic and electronic calculations were performed using density functional theory (DFT) calculations based on the local density approximation (LDA)[53] for the exchange-correlation functional, via the DMol$^3$ code.[54] The transmission Kikuchi diffraction was carried out in a Zeiss Auriga scanning electron microscope (SEM), which enabled us to quantitatively describe the crystal phase formed at each point of the micrograph with a minimum resolution of about 1 nm. The orientation of the crystals at each point was determined by a HKL Nordlys 2 system. Needle-shaped atom probe specimens with tip radius of < 100 nm were prepared by focused ion beam (FIB) on dual beam instruments. Pulsed-laser atom-probe tomography (APT) analyses were performed on a LEAP 4000X HR (Cameca Corp.), fitted with a reflectron in laser pulsed mode (355 nm, 5 ps, 150 pJ), at a base pressure of $3 \times 10^{-11}$ torr, a temperature of 24 K, and a detection rate of 2 ions per 1000 pulses. Local excess profiles were computed from conventional composition profiles as a function of the distance to the isoconcentration surfaces (proximity histograms) by comparing the concentration in each bin of the profile to the average concentration measured between 10 and 15 nm away from the position of the interface (0 in the profile).

**Conclusions**

In summary, we have demonstrated that doping GO into MgB$_2$ produces a reactive composite structure that significantly enhances $J_c$. This occurs via a combination of grain texturing and precipitation, providing both inter- and intra-granular flux pinning mechanisms. This approach to microstructural design has excellent prospects for more general applications. Moreover, the quantitative analysis and imaging of the low-dimensional nanostructures by APM has enabled a correlation to be established between the atomic-scale microstructure and the particular properties by guiding the inputs to (e.g.) DFT calculations. This offers a pathway towards a deeper understanding of fabrication processes relevant to electrochemical energy storage and optoelectronic devices containing low-dimensional nanostructured composites.

**Acknowledgements**

We thank S. H. Aboutalebi and K. Konstantinov for providing GO and P. W. Trimby for the technical help on transmission Kikuchi diffraction. The work was supported by the Australian Research Council, and we acknowledge scientific and technical support from the Australian Microscopy and Microanalysis Research Facility (AMMRF) at The University of Sydney.

Electronic Supplementary Information (ESI) available: [Crystallographic data on the proposed Mg(B,O)$_2$ phases, MgB$_2$O$_{0.25}$ and MgB$_2$O$_{0.5}$. Evidence of phase separation of Mg(B,O)$_2$ into nanocrystalline MgO and MgB$_2$, and evidence of Mg(B,O)$_2$ phase in X-ray diffraction with the X-ray index table for Mg(B,O)$_2$]. See DOI: 10.1039/b000000x/

1   J. Wei and D. Natelson, *Nanoscale*, 2011, **3**, 3509.
2   M. Fratini, N. Poccia, A. Ricci, G. Campi, M. Burghammer, G. Aeppli, A. Bianconi, *Nature*, 2010, **466**, 841.
3   S. R. Foltyn, L. Civale, J. L. MacManus-Driscoll, Q. X. Jia, B. Maiorov, H. Wang, M. Maley, M. *Nat. Mater.*, 2007, **6**, 631.
4   J. Weiss, C. Tarantini, J. Jiang, F. Kametani, A. A. Polyanskii, D. C. Larbalestier, E. E. Hellstrom, *Nat. Mater.*, 2012, **11**, 682.
5   C. B. Eom, M. K. Lee, J. H. Choi, L. J. Belenky, X. Song, L. D. Cooley, M. T. Naus, S. Patnaik, J. Jiang, M. Rikel, *Nature*, 2001, **411**, 558.
6   X. Song, G. Daniels, D. M. Feldmann, A. Gurevich, D. Larbalestier, *Nat. Mater.,* 2005, **4**, 470.
7   J. L. MacManus-Driscoll, S. R. Foltyn, Q. X. Jia, H. Wang, A. Serquis, L. Civale, B. Maiorov, M. E. Hawley, M. P. Maley, D. E. Peterson, *Nat. Mater.*, 2004, **3**, 439.
8   Z. Q. Ma, Y. C. Liu and Q. Cai, *Nanoscale*, 2012,**4**, 2060.
9   S. X. Dou, S. Soltanian, J. Horvat, X. L. Wang, S. H. Zhou, M. Ionescu, H. K. Liu, P. Munroe, M. Tomsic, *Appl. Phys. Lett*. 2002, **81**, 3419.
10  J. H. Kim, S. Zhou, M. S. A. Hossain, A. V. Pan, S. X. Dou, *Appl. Phys. Lett*. 2006, **89**, 142505-1-3.
11  Y. W. Ma, X. P. Zhang, G. Nishijima, K. Watanabe, S. Awaji, X. D. Bai *Appl. Phys.*, *Lett.* 2006, **88**, 072502.
12  X. P. Zhang, Y. W. Ma, Z. S. Gao, D. L. Wang, L. Wang, W. Liu, and C.R. Wang, *J. of Appl. Phys*., 2008, **103**, 103915.
13  S. X. Dou, W. K. Yeoh, J. Horvat, M. Ionescu, *Appl. Phys. Lett.*, 2003, **83**, 4996.
14  X. Xu, S. X. Dou, X. L. Wang, J. H. Kim, J. A. Stride, M. Choucair, W. K. Yeoh, R. K. Zheng, S. P. Ringer, *Supercond. Sci. Technol.*, 2010, **23**, 085003-1-5.
15  J .M. Rowell, *Supercond. Sci. Technol.*, **16**, R17 (2003).
16  X. Z. Liao, A. Serquis Y. T. Zhu J Y. Huang, D. E. Peterson F. M. Mueller, H. F. Xu, *Appl. Phys. Lett*. 2002, **80**, 4398; X. Z. Liao, A. Serquis Y. T. Zhu, J. Y. Huang, L. Civale, D. E. Peterson, F. M. Mueller, H. F. Xu, *J. Appl. Phys*. 2003, **93**, 6208.
17  R. F. Klie, J. C. Idrobo, N. D. Browning, A. Serquis, Y. T. Zhu, X. Z. Liao, F. M. Mueller, *Appl. Phys. Lett*., 2002, **80**, 3970; R. F. Klie, J. C. Idrobo, N. D. Browning, K. A. Regan, N. S. Rogado, R. J. Cava, *Appl. Phys. Lett.,* 2001, **79**, 1837.
18  A. P. Cocco, G. J. Nelson, W. M. Harris, A. Nakajo, T. D. Myles, A. M. Kiss, J. J. Lombardo and W. K. S. Chiu, *Phys. Chem. Chem. Phys*., 2013, **15**, 16377.
19  Y. Shimizu, H. Takamizawa, K. Inoue, F. Yano, Y. Nagai, L. Lamagna, G. Mazzeo, M. Perego and E. Prati *Nanoscale*, 2014, **6**, 706.
20  G. Scappucci, G. Capellini, W. M. Klesseac and M. Y. Simmons *Nanoscale*, 2013, **5**, 2600.
21  D. E. Perea, E. R. Hemesath, E. J. Schwalbac, J. L. Lensch-Falk, P. W. Voorhees, L. J. Lauho, *Nat. Nanotech*., 2009, **4**, 315.
22  K. Tedsree, T. Li, S. Jones, C. W. A. Chan, K. M. K. Yu, P. A. J. Bagot, E. A. Marquis, G. D. W. Smith, S. C. E. Tsang, Nanocatalyst. *Nat Nanotech*., 2011, **6**, 302.
23  L. M. Gordon, D. Joester, *Nature*, 2011, **469**, 194.
24  W. K. Yeoh, B. Gault, X. Y. Cui, C. Zhu, M. P. Moody, L. Li, R. K. Zheng, W. X. Li, X. L. Wang, S. X. Dou, G. L. Sun, C. T. Lin, S. P. Ringer, *Phys. Rev. Lett.*, 2011, **106**, 247002-1-4.
25  K. S. B. De Silva, S. H.  Aboutalebi, X. Xu, X. L. Wang, W. X. Li, K. Konstantinov, S. X. Dou, *Scripta Mater.*, 2013, **69**, 437.
26  W. K. Yeoh, S. X. Dou, *Physica C*, 2007, **456**, 170.
27  R. H. T. Wilke, S. L. Bud'ko, P. C. Canfield, D. K. Finnemore, R. J. Suplinskas, S. T. Hannahs, *Phys. Rev. Lett.* 2004, **92**, 217003-1-4.
28  W. X. Li, X. Xu, Q. H. Chen, Y. Zhang, S. H. Zhou, R. Zeng, S. X. Dou, *Acta Mater.*, 2011, **59**, 7268.
29  F. Tang, B. Gault S. P. Ringer, P. Martin, A. Bendavid, J. M. Cairney, *Scripta Materialia*, 2010, **63**, 192.
30  P. W. Trimby, *Ultramicroscopy*, 2012, **120**, 16.
31  S. X. Dou, W. K. Yeoh, O. Shcherbakova, D. Wexler, Y. Li, Z. M. Ren, P. Munroe, S. K. Chen, K. S. Tan, B. A. Glowacki, J. L. MacManus-Driscoll, *Adv. Mater*., 2006, **18**, 785.
32  A. Bagri, C. Mattevi, M. Acik, Y. J. Chabal, M. Chhowalla, V. B. Shenoy, *Nat. Chem*., 2010, **2**, 581.
33  R. K. Singh, Y. Shen, R. Gandikota, C. Carvalho, J. M. Rowell, N. Newman, *Appl. Phys. Lett*., 2008, **93**, 242504-1-3.
34  X. Y. Cui, J. E. Medvedeva, B. Delley. A. J. Freeman, C. Stampf, *Phys. Rev. B*, 2007, **75**, 155205-1-13.

35  Y. Yan, M. M. Al-Jassim, *Phys. Rev. B*, 2003, **67**, 212503-1-4.
36  S. Li, O. Prabhakar, T. T. Tan, C. Q. Sun, X. L. Wang, S. Soltanian, J. Horvat, S. X. Dou, *Appl. Phys. Lett.*, 2002, **81**, 874.
37  X. Y. Song, V. Braccini, D. C. Larbalestier, *J. Mater. Res.*, 2004, **19**, 2245.
38  X. L. Dong, Z. F. Dong, B. R. Zhao, Z. X. Zhao, X. F. Duan, L. M. Peng, W. W. Huang, B. Xu, Y. Z. Zhang, S. Q. Guo, L. H. Zhao, L. Li, *Phys. Rev. Lett.*, 1998, **80**, 2701.
39  J. M. Tranquada, Y. Kong, J. E. Lorenzo, D. J. Buttrey, D. E. Rice, V. Sachan, *Phys. Rev. B*, 1994, *50*, 6340-6351.
40  E. Arras, F. Lançon, I. Slipukhina, É. Prestat, M. Rovezzi, S. Tardif, A. Titov, P. Bayle-Guillemaud, F. D'Acapito A. Barski, *Phys. Rev. B*, 2012, **85**, 115204-1-10.
41  S. J. Park, J. An, J. W. Suk, R. S. Ruoff, *Small*, 2009, **6**, 210.
42  K. A. Yates, Z. Lockman, A. Kursumovic, G. Burnell, N. A. Stelmashenko, J. L. MacManus Driscoll, M. G. Blamire, *Appl. Phys. Lett.*, 2005, **86**, 022502-1-3.
43  T. Wenzel, K. G. Nickel, J. Glaser, H. J. Meyer, D. Eyidi, O. Eibl, *Phys. Stat. Sol. (a)*, 2003, **198**, 374.
44  C. G. Navarro, R. T. Weitz, A. M. Bittner, M. Scolari, A. Mews, M. Burghard, K. Kern, *Nano Lett.*, 2007, **7**, 3499.
45  M. Eisterer, C. Krutzler, H. W. Weber, *J. Appl. Phys.*, 2005, **98**, 033906-1.
46  G. Z. Li, M. D. Sumption, M. A. Susner, Y. Yang, K. M. Reddy, M. A. Rindfleisch, M. J. Tomsic, C. J. Thong, E. W. Collings, *Supercond. Sci. Technol.*, 2012, **25**, 115023-1-8.
47  S. K. Chen, M. Wei, J. L. MacManus-Driscoll, *Appl. Phys. Lett.*, 2006, **88**, 192512-1-3.
48  J. Wang, Y. Bugoslavsky, A. Berenov, L. Cowey, A. D. Caplin, L. F. Cohen, J. L. MacManus-Driscoll, L. D. Cooley, X. Song, D. C. Larbalestier, *Appl. Phys. Lett.*, 2002, **81**, 2026.
49  C. Cheng, Y. Zhao, *Appl. Phys. Lett.*, 2006, **89**, 252501-1-3.
50  X. L. Wang, S. X. Dou, M. S. A. Hossain, Z. X. Cheng, X. Z. Liao, S. R. Ghorbani, Q. W. Yao, J. H. Kim, T. Silver. *Phys. Rev. B*, **2010**, *81*, 224514-1-6.
51  Y. Yan, M. M. Al-Jassim, *J. Appl. Phys.*, 2002, **92**, 7687.
52  A. K. Bengtson, C. W. Bark, J. Giencke, W. Dai, X. X. Xi, C. B. Eom, D. Morgan, *J. Appl. Phys.*, 2010, **107**, 023902-1-4.
53  J. P. Perdew, Y. Wang, *Phys. Rev. B*, 1992, **45**, 13244.
54  B. Delley, *J. Chem. Phys.*, 1990, **92**, 508; *J. Chem. Phys.*, 2000, **113**, 7756.

**Figure 1** (a) Schematic diagram of the diffusion method used in the present study. B and GO were first mixed and pressed into a bar shape before infiltration with Mg powder at 800°C for 10 hours. (b) $J_c$ vs. H dependence obtained for bulk $MgB_2$ with nominal 0.5 wt.% of carbon doping level for graphene and graphene oxide doping at 5 and 20 K. A control sample of undoped $MgB_2$ was also prepared for comparison. GO-doping results in better $J_c$ over a wide range of applied magnetic field. (c) Temperature dependence of normalized $H_{irr}$ and $H_{c2}$ for different concentrations of GO. The normalized $H_{irr}$ is improved gradually with increasing doping level while the normalized $H_{c2}$ is almost unchanged beyond 1 wt.% GO

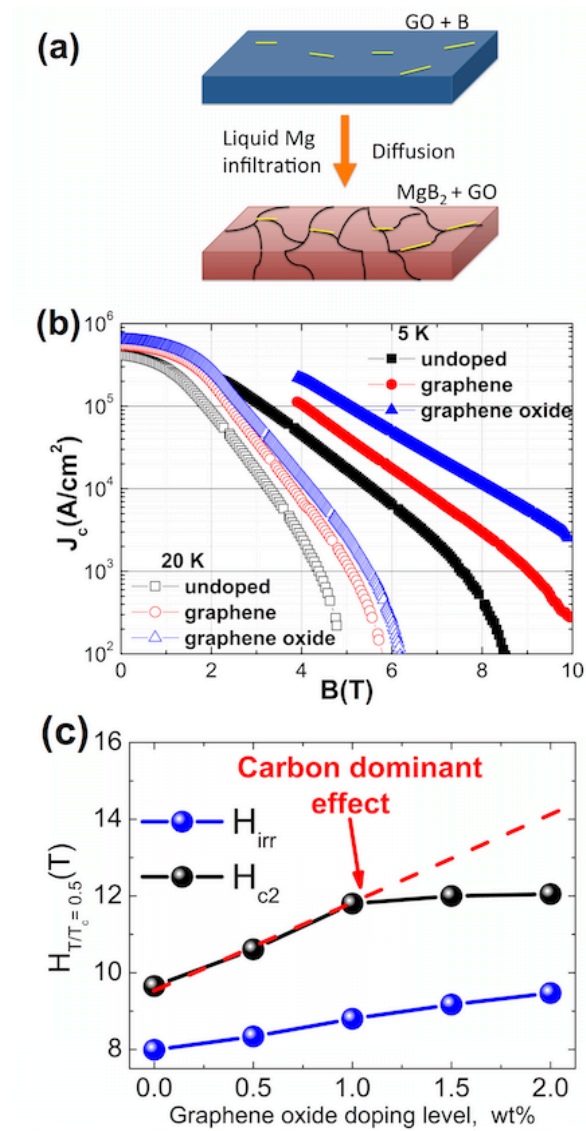

**Figure 2**. (a) 3D atom probe tomography reconstruction of GO-doped $MgB_2$. Each dot represents an individual atom of magnesium, boron, carbon, or oxygen, and as is readily visible in the close-ups shown in the regions of interest (ROI) 1 and 2, isoconcentration surfaces highlight the location of plate-shaped carbon and oxygen rich areas corresponding to reduced graphene oxide sheets. ROI 3 highlights the agglomeration of C and O at the grain boundary that crosses the dataset. (b) Local excess profiles extracted from the composition profiles around the isoconcentration surfacesexhibits significant differences in the carbon-to-oxygen ratio, which indicates that carbon and oxygen can be removed from GO sheets.

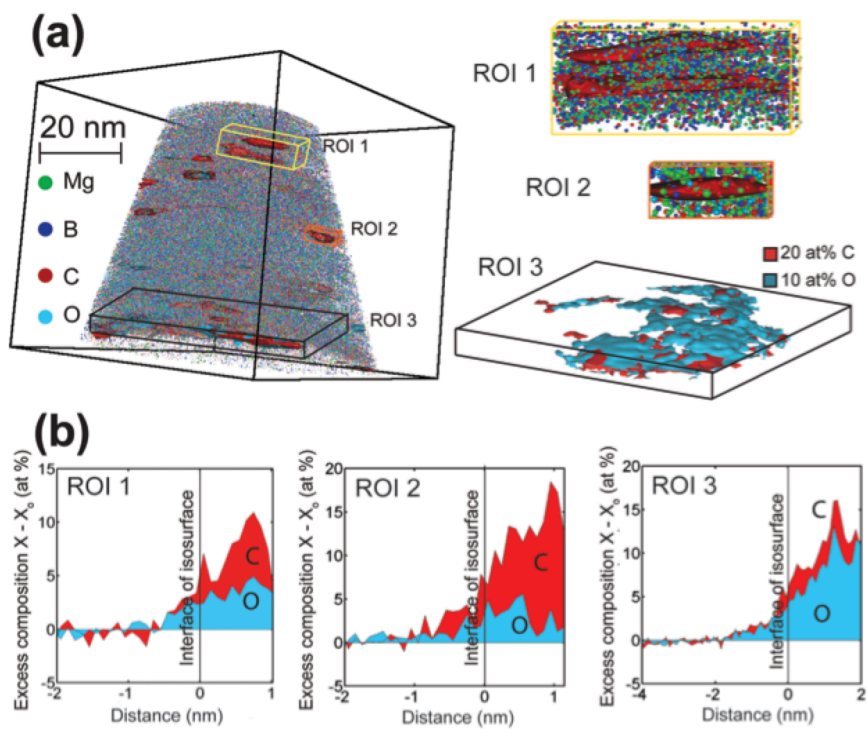

**Figure 3.** Microstructure characterization of the GO doped $MgB_2$ was performed by tranmission Kikuchi diffraction within a scanning electron microscope. (a) Overlapped diagram of the microstructure and grain orientation map, where the grains are coloured according to the inverse pole figure shown in the inset. Black lines highlight high-angle grain boundaries (> 10° misorientation); red lines highlight low-angle grain boundaries (5-10° of misorientation). Dark grey regions within grains correspond to grain or subgrain boundaries of super-low misorientation (< 5°), as shown in (c), where the misorientation along the A-B line remains below 5°. (b) Contoured pole figure sets along the {001} planes, displaying the texture of the GO-doped sample as compared to the undoped sample. The doped sample has only one texture orientation (I) compared to two texture orientations (I and II) in the undoped sample. (c) Misorientation profile along the line marked in (a).

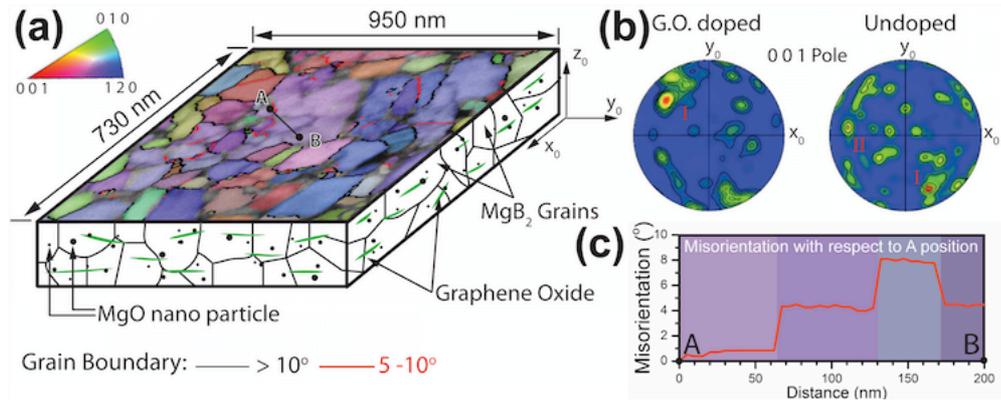

**Figure 4**. (a) Potential oxygen interstitial structures with single-layer three-oxygen clusters ($MgB_2O_{0.25}$) or double-layer six-oxygen clusters ($MgB_2O_{0.5}$). The oxygen interstitial modulated structure can be detected by TEM, as shown in (b), where nanoscale particles such as MgO particles surround the oxygen interstitial structure.

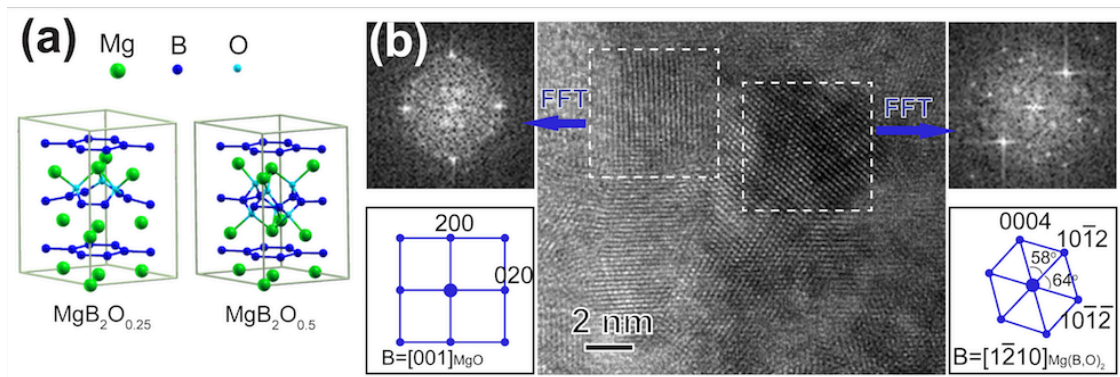

**Figure 5**. Dependence of intergranular and intragranular connectivity with GO concentration in MgB$_2$ samples. The monotonic increase in intragranular resistivity indicates an increased scattering from oxygen-containing defects within the grain. The decrease in the intergranular connectivity with GO concentration reflects an improvement in the grain connectivity, presumably as a result of the reduction of GO at the grain boundaries.

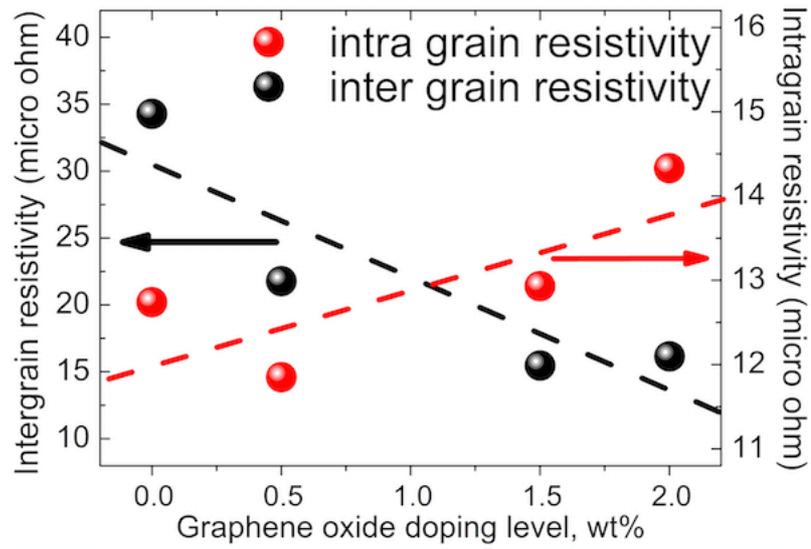